\documentclass[aps,pra,floatfix,twocolumn,superscriptaddress,showpacs]{revtex4}
\usepackage{bm}
\usepackage{epsf}
\usepackage{amssymb}
\usepackage{amsmath}
\usepackage{graphicx}
\usepackage{rotating}
\usepackage{epsfig}
\usepackage{psfrag}
\usepackage{amsmath}
\usepackage{hyperref}
\usepackage{subfigure}

\begin{document}

\def\bP{{\bf P}}
\def\bK{{\bf K}}
\def\bz{{\bf z}}
\def\bv{{\bf v}}
\def\br{{\bf r}}
\def\bq{{\bf q}}
\def\bp{{\bf p}}
\def\bQ{{\bf Q}}
\def\beq{\begin{equation}}
\def\eeq{\end{equation}}
\def\bea{\begin{eqnarray}}
\def\eea{\end{eqnarray}}
\def\bdm{\begin{displaymath}}
\def\edm{\end{displaymath}}
\def\e{\epsilon}
\def\ubs{|\mathrm{ub}\rangle}
\def\lbs{|\mathrm{lb}\rangle}
\def\ub{\mathrm{ub}}
\def\lb{\mathrm{lb}}
\def\ins{|\mathrm{i}\rangle}
\def\fis{|\mathrm{f}\rangle}
\def\in{\mathrm{i}}
\def\bnab{\boldsymbol{\nabla}}
\def\z0{\bar{z}_{\sigma}}
\def\dz{\delta z_{\sigma}}
\def\v0{\bar{\mathbf{v}}_{\sigma}}
\def\nup{n^{u}}
\def\nlo{n^{l}}

\title{Colliding clouds of strongly interacting spin-polarized fermions}
\author{Edward Taylor, Shizhong Zhang, William Schneider and Mohit Randeria}
\affiliation{Department of Physics, The Ohio State University, Columbus, Ohio, 43210}

\date{January 18, 2012}

\begin{abstract}
Motivated by a recent experiment at MIT, we consider the collision of two clouds of spin-polarized atomic Fermi gases close to a Feshbach resonance. We explain why two dilute gas clouds, with underlying attractive interactions
between its constituents, bounce off each other in the strongly interacting regime. Our hydrodynamic analysis,
in excellent agreement with experiment, gives strong evidence for a novel metastable many-body state with effective repulsive interactions.
\end{abstract}


\maketitle

\section{Introduction}
Ultracold atomic gases~\cite{Bloch08,Giorgini08}
open up new frontiers in the study of non-equilibrium dynamics of strongly interacting quantum systems.  
In contrast to electronic systems, the Fermi energy in quantum gases is of the order of a few kilohertz, so that questions about metastability and equilibration in quantum systems 
can now be explored in real time in the laboratory.  Exploring these new regimes and developing new theoretical tools to gain insight into the non-equilibrium dynamics
of many-body systems is an exciting challenge at the intersection of condensed matter and atomic-molecular-optical physics.

\begin{center}
\begin{figure}
\includegraphics[width=0.45 \textwidth]{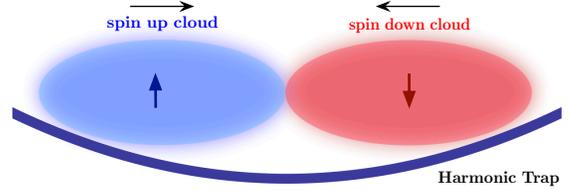}
\caption{Schematic of the colliding cloud experiment. }
\label{schematicfig}
\end{figure}
\end{center}

The recent experiment of Ref.~\cite{Sommer11} offers a beautiful example of non-equilibrium dynamics that is
completely unexpected and, at first sight, counterintuitive.
Two clouds of ultracold fermions prepared in different hyperfine-Zeeman (``spin") states 
$\sigma = {\uparrow}$ and ${\downarrow}$ are initially separated using a Stern-Gerlach field; see Fig.~\ref{schematicfig}. 
Once the field is turned off, the clouds are impelled together by the confining harmonic potential and they collide.
In the strongly interacting unitary regime these very dilute clouds are observed to bounce off each other, almost as if they were colliding billiard balls!  
This is truly remarkable in view of the fact
that the underlying atomic interaction is \emph{attractive}, and the 
equilibrium ground state is known to be a paired superfluid. 
Why then do the clouds behave as though the interactions are repulsive?

Two additional aspects of Ref.~\cite{Sommer11} are also noteworthy. 
Once the short-time bounce is damped out,
the centers of mass of the clouds remain separated for a considerable duration
at intermediate times ($\sim 100$ ms), before eventually merging. In fact, the analysis in Ref.~\cite{Sommer11} focused on precisely the slow merging of the clouds, or spin diffusion, at long times ($\sim 0.5$ s). 
Finally, the observed dynamics changes qualitatively as one moves out of the strongly interacting regime, 
with the clouds merging rapidly for weak interactions.

In this paper we gain insight into three aspects of this remarkable dynamics
using a hydrodynamic approach in the strongly interacting regime.
(1) We explain why the clouds repeatedly bounce off each other at short times.
(2) We also explain why the centers of mass of the two clouds remain separated at intermediate time scales.
We show that these results are natural consequences of a metastable many-body state
on the ``upper branch" of the Feshbach resonance where the effective interactions are repulsive~\cite{Pilati10,Chang11}.
(3) Finally, we provide insight into how the dynamics changes as a function of the 
strength of the interaction. 

\section{Hydrodynamics}
Our analysis is based on two hypotheses:
(i) The collisions between atoms are sufficiently rapid to establish local 
thermodynamic equilibrium when the clouds overlap. 
(ii) The loss from the upper branch scattering state is slow in the 
experiment of Ref.~\cite{Sommer11}.   
At unitarity, the two-body collision rate is very large, 
$1/\tau_2 \sim \epsilon_F$ (with $\hbar = k_B = 1$), 
for a range of temperatures $0.1 \lesssim T/\epsilon_F \lesssim 0.3$~\cite{Kinast05,Massignan05}. 
With a Fermi energy $\e_F \sim 10^4$Hz and an axial trap frequency $\omega_z \sim 10$Hz~\cite{Sommer11}, 
the gas in the overlap region of the clouds will reach local thermodynamic equilibrium 
within $\sim 10^{-3}$ trap periods.
Thus the clouds behave hydrodynamically in the overlap region~\cite{nonoverlap} 
and their dynamics reflect the (metastable) equation of state of the gas. 
(Hydrodynamics also describes well the
very different behavior observed in colliding clouds of {\it spin-balanced} superfluid gases at unitarity~\cite{Joseph11}.)

As the two clouds come together in the overlap region, the atoms are initially in scattering states, and the formation of two-body spin-singlet bound states
requires three-body collisions in order to satisfy kinematic constraints. The time scale
$\tau_3$ for such processes is not well understood for strongly interacting gases,
however, there are indications that it is enhanced close to unitarity~\cite{Zhang11}. 
In addition, the $\tau_3$ relevant to Ref.~\cite{Sommer11} is
further enhanced relative to the microscopic decay time
since three-body collisions are limited to a small overlap region for a fraction of the oscillation period $\omega_z^{-1}$. 

Now, for the experimentally relevant time interval $\tau_2 \ll t \ll \tau_3$, the rapid two-body collisions have already established thermal equilibrium, while the slow three-body recombination process has yet to drive the system into the lower branch. This allows us to model the cloud dynamics using Euler's equations
\beq 
\frac{\partial \bv_{\sigma}}{\partial t} + \frac{\bnab v^2_{\sigma}}{2}  = -\frac{\bnab}{m} \left(\frac{\partial {\cal{E}}}{\partial n_{\sigma}} + V_{\mathrm{trap}}\right) - \gamma \bv_{\sigma},\label{Euler}
\eeq
\beq 
\frac{\partial n_{\sigma}}{\partial t} + \bnab\cdot(n_{\sigma}\bv_{\sigma}) = 0,\label{continuity}
\eeq
with an appropriate energy density ${\cal{E}}[n_\sigma({\bf r})]$ for the scattering states 
(see below).  
Here $\bv_{\sigma}$ and $n_{\sigma}$ are the velocity and density of the $\sigma$ fermions, $V_{\mathrm{trap}} = m(\omega^2_{\perp}\rho^2 + \omega^2_z z^2)/2$ is the trap potential.
We include a phenomenological $\gamma$ to account for strong spin-current damping in the hydrodynamic regime~\cite{Vichi99,Bruun08}.  
Viscosity, describing the damping of in-phase current ($\bv_{\uparrow}=\bv_{\downarrow}$), is ignored since the motion of the colliding clouds is primarily out-of-phase~\cite{Massignan05,Cao11}.

As a result of tight confinement in the radial direction, the two modes with the lowest energies are the spin-dipole mode and axial breathing mode. Hence, we concentrate on the  centers-of-mass of the clouds and their widths along the $\hat{z}$-direction and reformulate (\ref{Euler}) and (\ref{continuity}) in terms of those variables. Using the notation $\langle \cdots\rangle_{\sigma}\equiv 1/N_{\sigma}\int d^3{\bf r} n_{\sigma}(\cdots)$, we define the center-of-mass positions 
$\z0 \equiv \langle z\rangle_{\sigma}$ and widths $\dz \equiv  \sqrt{8\langle (z-\z0)^2\rangle_{\sigma}}$.
The $\sqrt{8}$ ensures that $\dz$ coincides with the axial Thomas-Fermi (TF) radius $R_z$ 
in equilibrium.
(\ref{Euler}) and (\ref{continuity}) then lead to the exact equations of motion
\bea \ddot{\bar{z}}_{\sigma} + \omega^2_z \z0 &=& -\frac{1}{m}\left\langle \partial_z(\partial {\cal{E}}/\partial n_{\sigma})\right\rangle_{\sigma} -\langle \gamma 
v_{\sigma z}\rangle_{\sigma}\nonumber\\&&+\left\langle \bv_{\sigma}\cdot\bnab v_{\sigma z} - \partial_z v^2_{\sigma}/2\right\rangle_{\sigma}\label{zm}\eea
and
\bea \lefteqn{\ddot\dz + \omega^2_z\dz = 8\frac{\langle v_{\sigma z}^2 \rangle_{\sigma}}{\dz} - \frac{(\dot{\dz})^2}{\dz} - 8\frac{(\dot{\bar{z}}_{\sigma})^2}{\dz}}&&\nonumber\\&&\!\!\!\!-\frac{8}{m\dz}\left\langle (z-\z0) \partial_z(\partial {\cal{E}}/\partial n_{\sigma})\right\rangle_{\sigma} - \frac{8}{\dz}\langle (z-\z0)\gamma v_{\sigma z}\rangle_{\sigma}\nonumber\\&&\!\!\!\!+\frac{8}{\dz}\left\langle (z-\z0)[\bv_{\sigma}\cdot\bnab v_{\sigma z} - \partial_z v^2_{\sigma}/2]\right\rangle_{\sigma}. \label{dzm}\eea  

The problem can be further simplified while still retaining the essential physics of the (nonlinear) coupling between the spin-dipole and axial breathing modes by using a TF ansatz:
\beq n_{\sigma}(\br,t) =  \frac{R_z(2m\epsilon_F^0)^{3/2}}{6\pi^2\dz(t)}\Big[1\!-\!\Big(\frac{\rho}{R_{\perp}}\Big)^2\! -\! \Big(\frac{z-\z0(t)}{\dz(t)}\Big)^2\Big]^{3/2},\label{ansatz}\eeq
where $R_{\alpha}\equiv \sqrt{2\epsilon_F^0/m\omega^2_{\alpha}}$ is the TF radius along the $\alpha$-axis 
and $\epsilon_F^0 = (\omega^2_{\perp}\omega_z)^{1/3}(3N)^{1/3}$ 
is the chemical potential of an ideal two-component Fermi gas ($N_{\uparrow}=N_{\downarrow}=N/2$).  
This ansatz allows for a time-dependent center-of-mass $\z0$ as well as 
axial compression of the clouds.  
The continuity equation (\ref{continuity}) leads to the velocity field ${\bf v}_\sigma=v_\sigma\hat{z}$ with $v_{\sigma}(z,t) = \dot{\bar{z}}_{\sigma}- \z0\dot{\dz}/\dz + z\dot{\dz}/\dz$.  
Axial and radial symmetry lets us set $\bar{z}_{\uparrow}\equiv \bar{z} = -\bar{z}_{\downarrow}$ and $\delta z_{\uparrow} = \delta z_{\downarrow}\equiv \delta z$.  Using (\ref{ansatz}) in (\ref{zm}) and (\ref{dzm}) gives the coupled nonlinear integro-differential equations:
\beq
\left(\frac{d^2}{dt^2}+\omega^2_z\right)\bar{z}(t) = -\frac{1}{mN_{\uparrow}}\int d^3r \frac{\partial{\cal{E}}}{\partial n_{\uparrow}}\frac{\partial n_{\uparrow}}{\partial\bar{z}} - \langle \gamma v_{\uparrow}\rangle_{\uparrow}\label{barz},\eeq
and
\bea
\left(\frac{d^2}{dt^2}+\omega^2_z\right)\delta z(t) &=& -\frac{8}{mN_{\uparrow}}\int d^3r \frac{\partial{\cal{E}}}{\partial n_{\uparrow}}\frac{\partial n_{\uparrow}}{\partial\delta z}\nonumber\\&& - \frac{8}{\delta z}\langle (z-\bar{z}(t))\gamma v_{\uparrow}\rangle_{\uparrow}.\label{deltaz}\eea

In the unitarity regime, the spin-current damping~\cite{Bruun08} assumes a particularly simple form when the relative velocity is of order of the Fermi velocity, as is the case at early times for two clouds initially separated by $R_z$. In the overlap region, we take
\beq \gamma = \widetilde{\gamma}\sqrt{\epsilon_{F\uparrow}\epsilon_{F\downarrow}},\eeq where 
$\widetilde{\gamma}$ is of order unity at unitarity and 
$\epsilon_{F\sigma}({\bf r})=(6\pi^2 n_{\sigma}(\br,t))^{2/3}/2m$ is the local Fermi energy. 
This simple choice of damping is, if anything, an overestimate, and
we have checked that our results are robust against reasonable modification of the damping parameter.

\begin{center}
\begin{figure}
\includegraphics[width=0.45 \textwidth]{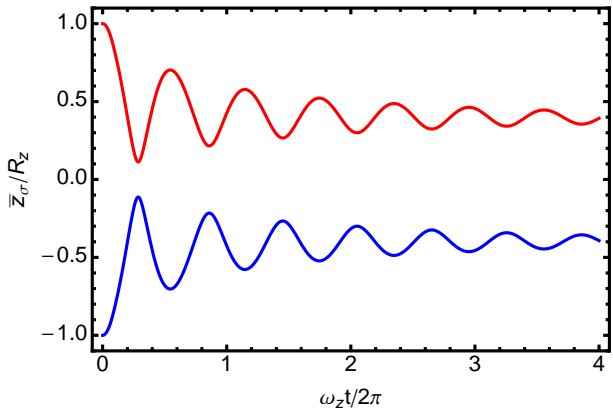}
\caption{(Color online) Time dependence of the center-of-mass positions $\z0$ of the two spin clouds at unitarity,
with the red (upper) and blue (lower) curves denoting the two spin species.
We use the upper branch energy functional (see text), with $\epsilon_F^0$ 
corresponding to $N=7.5\times 10^5$ atoms, trap anisotropy $\omega_{\perp}/\omega_z=10$,
and damping $\widetilde{\gamma}=1$.  
Initially the two clouds are displaced from the trap center by the axial Thomas-Fermi radius, $\bar{z}_{\uparrow}(t=0)=R_z, \bar{z}_{\downarrow}(t=0)=-R_z$ 
with initial axial width $\delta z(0)=R_z$ equal to its equilibrium value.  
}
\label{upperbouncefig}
\end{figure}
\end{center}

\section{Energy Functional}
We now need to specify the energy functional ${\cal E}[n_\sigma]$ relevant 
for dynamics at times $\tau_2 \ll t \ll \tau_3$. 
The ground state (``lower branch'' of the Feshbach resonance) for 
scattering length $a_s > 0$ necessarily involves bound pairs. But for $t \ll \tau_3$, the
three-body processes required to relax to this state have not yet occurred.  
The system has, however, developed 
two-body correlations characteristic of the metastable ``upper branch" state,
studied theoretically in Refs.~\cite{LeBlanc09,Pilati10,Chang11},
motivated by an earlier experiment~\cite{Jo09}. 
The approximate many-body wavefunction in the upper branch is 
\beq
\Psi=\Big[\prod_{i,j}f(\vert\br_{i\uparrow}-\br_{j\downarrow}\vert)\Big]\Phi_S(\{ \br_{i\uparrow}\})\Phi_S(\{\br_{j\downarrow}\}),
\label{jastrow}
\eeq
where $\Phi_S(\{ \br_{i\sigma}\})$'s are Slater determinants and
the Jastrow factor $f(r)$ describes the short-range correlations between fermions.   
The effective repulsion between $\uparrow$ and $\downarrow$ atoms
in the upper branch is crucially related to the node
in  $f(r)$, in contrast to the nodeless Jastrow factor for the lower branch.  
For small $a_s>0$, the node occurs at $a_s$,
with $f(r)\sim (1-a_s/r)$ similar to the two-body problem.  For large $a_s$, 
the node saturates~\cite{Chang11} to $\sim 1/k_F$, the only length scale at unitarity.
Quantum Monte Carlo (QMC) studies~\cite{Pilati10,Chang11} of the upper-branch wavefunction (\ref{jastrow}) reveal that the system
undergoes (ferromagnetic) phase separation for sufficiently strong interactions  
$k_Fa_s\gtrsim 1$.  

We need ${\cal E}[n_\sigma]$ for arbitrary polarization, which has not been studied by QMC.
We thus use the lowest order constraint variational (LOCV) approximation~\cite{Pand7173},
which has been used for the upper branch~\cite{Heiselberg10} and is in close
agreement with QMC data~\cite{Pilati10,Chang11}.  As shown in the Appendix, the upper branch LOCV energy density is
\beq
{\cal{E}}=\frac{3}{5}\e_{F\uparrow}n_{\uparrow}+\frac{3}{5}\e_{F\downarrow}n_{\downarrow}+\frac{1}{2}(n_\uparrow \lambda_\downarrow+n_\downarrow \lambda_\uparrow).\label{ELOCV}
\eeq
The interaction energies $\lambda_\uparrow$ and $\lambda_\downarrow$ are functions of $k_Fa_s$ and $x = n_\downarrow/n_\uparrow$.  
Hydrodynamics requires $T\gtrsim 0.1\epsilon_F$~\cite{Massignan05}; for simplicity, we use the zero temperature LOCV energy functional to study the dynamics, expecting it to qualitatively describe the physics at low temperatures. It is important to note here that while we rely on the approximate LOCV calculation to obtain the equation of state, the underlying physics of the effective repulsive interaction is independent of the particular scheme used and a more precise calculation would only leads to quantitative improvement without changing the crucial physical picture of the bounce.
 
\section{Dynamics at unitarity}
We solve (\ref{barz}) and (\ref{deltaz}), using (\ref{ELOCV}) and (\ref{ansatz}); for details see Appendix~\ref{Appendixnumerical}. 
The results at unitarity are shown in Fig.~\ref{upperbouncefig}.  The two clouds bounce off each other for several oscillations due to the repulsive nature of the upper branch functional.  The period of the initial bounce is roughly $0.56(2\pi/\omega_z)$, slightly less than the experimental value $0.61(2\pi/\omega_z)$~\cite{Sommer11}.  We attribute this difference to the simple ansatz (\ref{ansatz}) used to model the dynamics.  
The bounce persists for several cycles in spite of the very large damping ($\epsilon_F^0/\omega_z\gg 1$)  because the overlap between the two clouds is small.  Once the oscillation is damped out, the 
centers of mass of the two clouds remain segregated, with a final $(\bar{z}_{\uparrow} - \bar{z}_{\downarrow}) \simeq 0.4$ times the initial separation.
This intermediate time behavior reflects the tendency for system to phase segregate in the upper branch at unitarity. 

\begin{center}
\begin{figure}
\includegraphics[width=0.45 \textwidth]{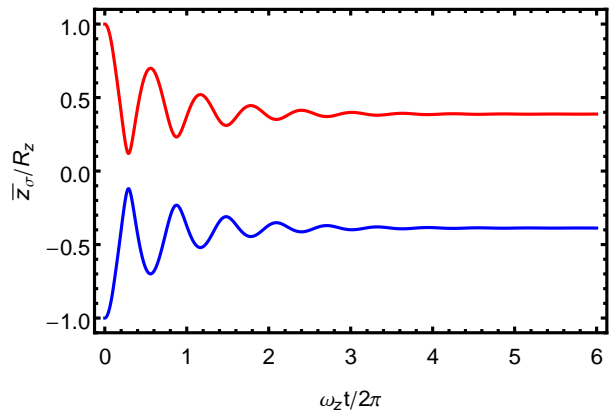}
\caption{(Color online) Time dependence of the center-of-mass positions $\z0$ of the two spin clouds for $k_F(0)a_s=2$. The parameters are the same as in Fig.\ref{upperbouncefig}.}
\label{bounce2fig}
\end{figure}
\end{center}

\section{Collisional dynamics away from unitarity}
Away from unitarity, it is harder to reach the hydrodynamic regime.  If $\omega_z$ is sufficiently small, however, it will always be the case that the dynamics in this direction are hydrodynamic.  The greater challenge at finite $a_s > 0$ is that $\tau_3$ may not be much greater than $\tau_2$.  As $a_s$ decreases from unitarity, one expects
that $\tau_3$ reaches a minimum for $k_F a_s\sim 1$ and then becomes large again~\cite{Petrov03}, 
with $\tau_3\sim (na^3_s)^{-2}\epsilon^{-1}_F$ for $k_Fa_s \ll 1$.  In contrast, 
the cross-section $a^2_s$ determines $\tau_2\sim (na^3_s)^{-2/3}\epsilon^{-1}_F$.  
Hence, for small $k_Fa_s$, again $\tau_3\gg \tau_2$.  However, when $k_Fa_s\sim 1$, it is conceivable that $\tau_3\sim \tau_2$.

\begin{center}
\begin{figure}
\includegraphics[width=0.45 \textwidth]{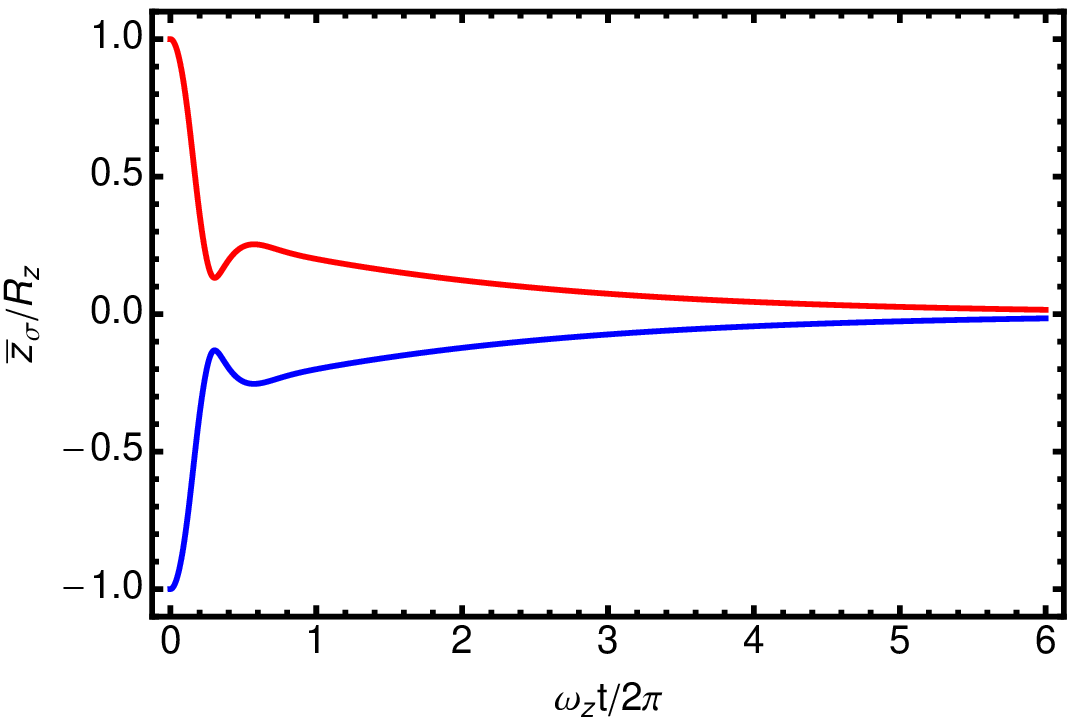}
\caption{(Color online) Time dependence of the center-of-mass positions $\z0$ of the two spin clouds for $k_F(0)a_s=0.5$. The parameters are the same as in Fig.\ref{upperbouncefig}. }
\label{bounce3fig}
\end{figure}
\end{center}

With these caveats in mind, we solve (see Appendix~\ref{Appendixnumerical} for details)
our hydrodynamic equations away from unitarity 
to understand the relationship between the intermediate-time dynamics, 
after the bounce is damped out, and the upper branch equation of state:
Fermi liquid ($k_Fa_s\lesssim 1$) versus phase separated ($k_Fa_s\gtrsim 1$).
The solutions of  (\ref{barz}) and (\ref{deltaz}) at $k_F(0)a_s=2$ and $0.5$ (using the same
initial conditions and $\epsilon_F^0$ as in Fig.~\ref{upperbouncefig}), are shown in 
Figs.~\ref{bounce2fig} and  \ref{bounce3fig}.  
Here $k_F(0) = \sqrt{2m\epsilon_F^0}$ is the ideal gas Fermi wavevector at the cloud center.
Appropriate for the smaller interaction strength, we use an expression for the spin-current damping obtained from  
kinetic theory~\cite{Vichi99}.  When symmetrized in the spin components, it takes the simple form \cite{comment1}
\beq \gamma=(4/9\pi)(\sqrt{k_{F\uparrow}k_{F\downarrow}}a_s)^2
\sqrt{\epsilon_{F\uparrow}\epsilon_{F\downarrow}}.\eeq

Even at small $k_F(0)a_s$, the clouds exhibit a weak bounce due to compressional recoil, 
which damps out very quickly. We see, however, a clear difference between Fig.~\ref{bounce2fig} ($k_Fa_s\gtrsim 1$), where the  
centers of mass remain separated, and Fig.~\ref{bounce3fig} ($k_Fa_s\lesssim 1$), where they merge. 
This behavior is qualitatively similar to the experimental results shown
in Supplementary Fig.~1 of Ref.~\cite{Sommer11}.

\section{Discussion}
We now compare our results in detail with experiments.
As seen from Fig.~\ref{upperbouncefig}, our results at unitarity -- the short-time bounce, the damping of the oscillations and the separation of the clouds at intermediate times -- are all
in very good agreement with Ref.~\cite{Sommer11}.  Away from unitarity, at finite values of $k_F a_s > 0$,
our results are qualitatively similar to the data
in Supplementary Fig.~1 of Ref.~\cite{Sommer11}.
For times $ \omega_z^{-1}\ll t \ll \tau_3$, say, $t=200$ ms, 
the system remains phase segregated for large $k_Fa_s$ [Figs.~1(f,g)],
while completely mixed for small $k_Fa_s$ [Figs.~1(c,d,e)]. 
In fact, a very rough estimate for the critical $k_F a_s$ can be read off from the 
experimental data: it is between $0.26$ [Fig.~1(e)] and $1.2$ [Fig.~1(f)].
The long time behavior, not described here, will be dominated by spin diffusion~\cite{Sommer11}
at non-zero temperatures and three-body processes.  

Finally, we comment on the experiment of Jo {\it et al.}~\cite{Jo09} 
in which a spin-balanced mixture, initially at a small $a_s>0$, is swept close to resonance, 
generating strong interactions in the upper branch. 
It appears, however, that rapid losses render it unstable to 
the lower branch within a very short time ($\tau_{3,\mathrm{micro}}\sim 1$ms)
\cite{Zhang11,Pekker11,Sanner11} and phase separated ferromagnetic domains
are not observed.

In contrast, the specific initial configuration in the 
colliding cloud experiment~\cite{Sommer11} proves to be crucial
for the metastability of the upper branch.
Three-body loss is limited spatially within the overlap region between different spins
at the trap center, and temporally to a fraction of the oscillation period. 
As such, the effective $\tau_3$ is greatly enhanced ($\tau_3\gg \tau_{3,\mathrm{micro}}$) and one can study the metastable upper branch.  Note that while $\tau_{3,\mathrm{micro}}$ is small, $\sim 10/\epsilon_F$~\cite{Sanner11}, it is still an order of magnitude greater than $\tau_2\sim \epsilon_F$, and thus local thermodynamic equilibrium can be established in the upper branch.  

In summary, we have investigated the short-time bounce dynamics of the recent MIT experiment~\cite{Sommer11}. The bounce frequency we found is in good agreement with the experiments. Furthermore, the experimental observation that the two clouds sit side-by-side 
for as long as $\sim 100$ms validates our assumption of a long $\tau_3$. 
We argue that the intermediate time behavior of the two colliding clouds indicates that the system favors a phase separated metastable state for larger value of $k_Fa_s$. We cannot see how a lower-branch energy functional, with attractive interactions forming bound pairs, could lead to the observed dynamics.

{\it Note added:} Recently, a Boltzmann approach was used to investigate the same problem in the high-$T$ regime~\cite{Goulko11}. The system is found to be strongly hydrodynamic for $k_Fa_s \gtrsim 1$.

\begin{acknowledgments}
We acknowledge discussions with S.~Stringari,
S-Y.~Chang, T.L.~Ho, N.~Trivedi and M.~Zwierlein.  
We gratefully acknowledge support from ARO W911NF-08-1-0338 (ET), NSF-DMR 0907366 (SZ, ET),
NSF-DMR 1006532 (WS, MR).
\end{acknowledgments}

\appendix

\section{Upper branch and LOCV}
\label{Appendix}
In this Appendix, we present a detailed discussion of  the so-called ``upper branch'' of a Feshbach scattering resonance in two-component Fermi gases~\cite{Pilati10,Chang11}
as well as the lowest order constraint variational (LOCV) method~\cite{Pand7173},
which we use to calculate the upper branch equation of the state~\cite{Heiselberg10}.

\subsection{Upper Branch}
The concept of the upper branch of a Feshbach resonance comes from the two-body problem
(either in a finite box or in a harmonic potential), where it is completely well-defined
for all values of $a_s$, positive or negative.
The two-body wavefunction in the upper branch is a scattering state
with a single node that makes it orthogonal to the ground state (lower branch) wave-function. 

This notion has been generalized to the many-body case~\cite{Pilati10,Chang11} by writing 
a Jastrow-Slater wavefunction, where the Jastrow
correlation factor has a node; see Eq.~(9) of the main paper and the discussion
immediately following this equation.

To summarize our definition of the upper branch in the many-body problem, 
any sensible definition must {\it at least} satisfy the following conditions \cite{Chang11}:\\
(1) The many-body wavefunction includes, apart from the nodes introduced by the Pauli principle, one additional node for any pair of fermions with opposite spin.\\
(2) The wavefunction should reduce, in the limit of the two-body problem, to that of the scattering states with one node in the relative wavefunction.\\
(3) The energy of the system must be larger than that of the non-interacting Fermi gas
and, it should reduce to the perturbative result in the weakly interacting regime $0 < k_F a_s \ll 1$.

We emphasize that orthogonality with the many-body ground state (of the BCS-BEC crossover) is \emph{not}
sufficient to be on the upper branch.

\subsection{LOCV}
LOCV is an approximation scheme to evaluate the energy of the Jastrow-Slater state
that originated in nuclear many-body physics~\cite{Pand7173} and has been used for
various quantum fluids.
Within LOCV, the energy of the Jastrow-Slater state [Eq.~(9) of paper]
is given by
\bea\label{ELOCV1}
{\cal{E}}=\frac{3}{5}\e_{F}n+\frac{n^2}{4}\!\!\int \!\!d^3r f^*(r)\!\left[-\frac{\nabla^2}{m}+v(r)\!\right]\!f(r),
\eea
where $v(r)$ is the short-range two-body potential, $n=n_\uparrow+n_\downarrow$ is the total density,
and $\e_F=(3\pi^2 n)/2m$ is the Fermi energy. 
The effect of a zero-range contact potential may be written in terms of  
the Bethe-Peierls boundary condition $\lim_{r\to 0}(rf(r))'/(rf(r))=-1/a_s$. 
In addition, within LOCV, the Jastrow function $f(r)$ satisfies the following conditions:  
$f(r\geq d)=1$ and $f'(d)=0$. The ``healing length" $d$ in turn is defined so that 
\beq
2 \pi n\int_0^{d} dr r^2 f^2(r)=1.
\label{cons}
\eeq
Variation of the energy (\ref{ELOCV1}) with respect to $f(r)$, while taking into account the constraint (\ref{cons}) by a Lagrange multiplier $\lambda$, gives us the ``Schr\"{o}dinger'' equation
\beq
\left[-\frac{\nabla^2}{m} + v(r)\right]f({\bf r})=\lambda f({\bf r}).
\eeq
Retaining only the $s$-wave part of this equation, we find the general solution with one node  
is given by
\beq
f(r)=\frac{d}{r}\frac{\sin(\kappa (r-b))}{\sin(\kappa (d-b))}
\eeq
where $\kappa=\sqrt{m\lambda}$. We find that $f(d)=1$ and $f'(d)=0$ lead to
$\kappa d =\tan (\kappa (d - b))$ and the Bethe-Peierls boundary condition gives
$\kappa a_s=\tan\kappa b$.
With the normalization condition (\ref{cons}), we can solve for $\kappa$, $d$ and $b$ for each value of scattering length $a_s$. The energy of the system is given simply by
\beq
{\cal{E}}=\frac{3}{5}\e_{F}n+\frac{1}{2}n\lambda.
\eeq

In the case of our interest, however, the system is not necessarily balanced, so that
$x\equiv n_{\downarrow}/n_{\uparrow}$ need not be unity.
We have to extend the LOCV calculation to the spin-imbalanced case in which the Slater determinants will have 
different sizes. We find the energy of the system is given by
\bea
{\cal{E}}&=&\frac{3}{5}\e_{F\uparrow}n_{\uparrow}(1+x^{\frac{5}{3}}) \\\nonumber
&+&n_\uparrow n_\downarrow\!\!\int \!\!d^3r f^*(r)\!\left[-\frac{\nabla^2}{m}+v(r)\!\right]\!f(r).
\eea
Now, in general, there is no unique way to enforce the normalization conditions as in (\ref{cons}). A natural extension is to use both normalizations
\beq
4 \pi (n-n_{\sigma})\int_0^{d_{\sigma}} dr r^2 f_{\sigma}^2(r)=1,\label{NORM}
\eeq
which introduces two healing lengths, $d_{\uparrow}$ and $d_{\downarrow}$. 
Accordingly, we shall introduce two Lagrange multipliers $\lambda_\uparrow$ and $\lambda_\downarrow$
and the energy of the system can then be written as
\beq
{\cal{E}}=\frac{3}{5}\e_{F\uparrow}n_{\uparrow}+\frac{3}{5}\e_{F\downarrow}n_{\downarrow}+\frac{1}{2}(n_\uparrow \lambda_\downarrow+n_\downarrow \lambda_\uparrow).\label{ELOCV2}
\eeq

\begin{widetext}
\begin{center}
\begin{figure}[h]
\includegraphics[width=1 \textwidth]{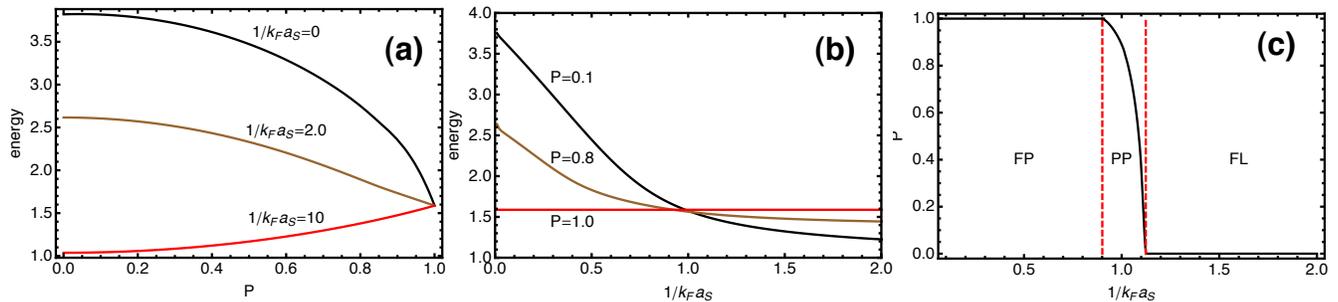}
\caption{The single particle energy in the upper branch, obtained from lowest order constraint variational method. The energy are in unit of $3/5\epsilon_F$  {\bf (a)}, energy as a function of the polarization $P$ for various values of $k_Fa_s$ as indicated in the figure. {\bf (b)}, energy as a function of the interaction parameter $k_Fa_s$ for various values of polarization $P$ as indicated in the figure. {\bf (c)}, phase diagram of the interacting two-component Fermi gases. The two vertical red dashed lines indicates the separation between three phases. For $1/(k_Fa_s)>1.124$, the system is in the Fermi liquid (FL) phase. For $1.124>1/(k_Fa_s)>0.91$, the system is in the partially polarized phase (PP) and for $1/(k_Fa_s)<0.91$, the system is fully polarized.}
\label{sup_fig}
\end{figure}
\end{center}
\end{widetext}

The above energy functional reduces to the usual perturbative result in the weak coupling limit and also, as we shall show later, reproduces the phase diagram of the interacting fermion system given by Monte Carlo methods \cite{Pilati10,Chang11} with very good agreement. We note that the same energy functional (\ref{ELOCV2}) was analyzed recently by Heiselberg \cite{Heiselberg10}, in the context of the 
experiment of ref.~\cite{Jo09}. 

In Fig.~\ref{sup_fig} {\bf (a)}, we show the upper branch energy (\ref{ELOCV2}) as a function of polarization $P \equiv (1-x)/(1+x)$ for various values of $k_Fa_s$. For small  $k_Fa_s$ (in fact, $k_Fa_s<0.91$), the minimum energy is attained at $P=0$. For larger values, $k_Fa_s\geq 1.124$, the minimum value is attained at $P=1$. For intermediate values of $k_Fa_s$, the minimum value occurs at a  value of $0<P_c<1$. In Fig.~\ref{sup_fig} {\bf (b)}, we show the upper branch energy as a function of $1/k_Fa_s$ for various values of the polarization $P$. Note that for $P=1$, {\it i.e.}, completely polarized, the energy is completely flat, since there is no interaction between the polarized fermions.

In Fig.~\ref{sup_fig} {\bf (c)}, we show the phase diagram of the system as obtained from LOCV. For $1/(k_Fa_s)>1.124$, the system is a homogeneous mixture of the two hyperfine-Zeeman states. We call this a Fermi liquid (FL) state. For $1.124>1/(k_Fa_s)>0.91$, the system is in the partially polarized phase (PP). Here the transition as predicted by the LOCV method is second order and accompanied by a divergent spin susceptibility \cite{Heiselberg10}. Note that the value of $1/(k_Fa_s)$ at the transition is very close to that predicted by Monte Carlo calculations \cite{Pilati10,Chang11} and compares favorably with the calculation in Ref.~\cite{Recati11}. Lastly, for $1/(k_Fa_s)<0.91$, the system is a fully polarized, non-interacting Fermi gas.

\section{Numerical details}
\label{Appendixnumerical} 
In this section, we explain how we use the LOCV energy
in the numerical solution of the hydrodynamic equations (6) and (7) of the main text. The important quantity that enters both equations is
\bea
\frac{\partial{\cal{E}}}{\partial n_{\uparrow}}=\e_{F\uparrow}+\frac{1}{2}\left(\lambda_\downarrow + n_\uparrow \frac{\partial \lambda_\downarrow}{\partial n_{\uparrow}} + n_\downarrow \frac{\partial \lambda_\uparrow}{\partial n_{\uparrow}}\right).
\eea
We express this in dimensionless form as 
\begin{widetext}
\bea
\widetilde{\mu}_\uparrow\equiv\frac{1}{\epsilon_{F\uparrow}}\frac{\partial{\cal{E}}}{\partial n_{\uparrow}}=1+\frac{1}{2}\left(\frac{5}{3}\widetilde{\lambda}_\downarrow -x \frac{\partial \widetilde{\lambda}_\downarrow}{\partial x}-\frac{1}{3}\xi \frac{\partial \widetilde{\lambda}_\downarrow}{\partial \xi}+\frac{2}{3}x\widetilde{\lambda}_\uparrow -\frac{1}{3}x\xi \frac{\partial \widetilde{\lambda}_\uparrow}{\partial \xi} \right).
\label{chem}
\eea
\end{widetext}
Here $\xi = 1/(k_F a_s)$ and all energies are scaled by $\epsilon_{F\uparrow}$, so that
$\widetilde{\lambda}_\uparrow \equiv {\lambda_\uparrow}/{\epsilon_{F\uparrow}}$ \textit{etc.} We note that at unitarity, $\xi=0$, and the expression involves only a single
derivative ${\partial \widetilde{\lambda}_\downarrow}/{\partial x}$ that has 
to be evaluated numerically.
For arbitrary $a_s$, however, the expression involves two additional numerical derivatives. 
To simplify the numerics at ``small'' $a_s$, we find that we can make
a polynomial fit to the equation (\ref{chem}) as a function of $x$ and $1/\xi$. 
we find that, for the values of $k_{F\uparrow}a_s \lesssim 2 $ we are considering, 
a good fit is provided by the expression
\bea
\widetilde{\mu}_\uparrow=1+x\left(\frac{3}{4\pi} k_{F\uparrow}a_s+
0.27 (k_{F\uparrow}a_s)^2\right)+\cdots
\eea
where the coefficient $0.27$ is very close to the value obtained by Galitskii~\cite{galitskii}, $\frac{4}{15\pi^2}(11-2\ln 2)\approx 0.259$.


\begin{thebibliography}{99}
\bibitem{nonoverlap}
While hydrodynamics is valid in the overlap region, the 
atoms in the non-overlap region are non-interacting. However, due to Pauli repulsion, 
the non-overlapping regions essentially execute a rigid-body dipole oscillation, 
insensitive~\cite{Vichi99} to the choice of a collisional (hydrodynamic) or collisionless description.
\bibitem{comment1} We choose $T\approx 0.2 T_F$ leading to $F(T)=1$ in eq.~(20) of Ref.~\cite{Vichi99}. Note that $F(T)$ is weakly $T$-dependent for $0.2 T_F< T < T_F$.
\end{thebibliography}
\end{document}